\documentclass[12pt,preprint]{aastex}
\usepackage{graphics}
\shorttitle{Imaging with GAPMs}
\shortauthors{Debes et al.}

\begin{document}
\title{High Contrast Imaging with Gaussian Aperture Pupil Masks}
\author{John H. Debes \& Jian Ge}
\affil{Pennsylvania State University, 525 Davey Lab, University Park, USA
}

\begin{abstract}
Gaussian aperture pupil masks (GAPMs) can in theory achieve the contrast
requisite for directly imaging an extrasolar planet around a nearby
solar type star.  We outline the process
 of designing, fabricating, and testing a GAPM
for use on current telescopes and specifically the Penn State near-IR Imager and Spectrograph (PIRIS) at the Mt. 
Wilson 100$^{\prime\prime}$ telescope.  We find that the initial prototype
observations are quite successful, achieving a contrast similar to a traditional Lyot coronagraph without blocking any light from a central object
and useful for finding faint companions to nearby young solar analogues.  In 
the lab we can reproduce the expected PSF reasonably well
and with a single aperture design which achieves $\sim4 \times\ 10^{-5}$ contrast at 10$\lambda/D$.
  We find that small inaccuracies in the mask fabrication process and 
insufficient correction of the atmosphere contribute the most degradation to 
contrast at these levels.   
\end{abstract}


\section{Introduction}
\label{intro}
The search to directly image an extrasolar planet around a nearby
solar type star requires contrast
levels of $\sim$10$^{-10}$ a few $\lambda / D$ from the central star, where 
$\lambda$ is the wavelength of light observed and $D$ is the diameter of 
the observing telescope's primary mirror.
Scattered light in a telescope
and the diffraction pattern of the telescope's aperture limit the contrast
possible for direct detection of faint companions \citep{brown91}.  The circular aperture of
telescopes creates a
sub-optimal diffraction pattern, the so-called Airy Pattern which is
azimuthally symmetric.  In addition, the intensity in the diffraction pattern
of the circular
aperture declines as $(\theta/\theta_o)^{-3}$ where $\theta_o=\lambda/D$. 
 Currently the best way to diminish
the Airy pattern is to use a coronagraph by using the
combination of a stop in the focal plane that rejects a majority of the
central bright object's light and a Lyot stop in the pupil plane to reject residual light diffracted to the edge \citep{lyot,malbet96,sivar01}.  Several recent ideas explore the use of alternative ``apodized'' apertures for high contrast
imaging in the optical or near-infrared \citep{nisenson01,spergel01,ge02,ks03}.
These designs revisit concepts first experimented with in the field of optics
\citep{jacquinot64}.  Other designs, such as the band
limited mask or notch-filter mask,
 seek to null the light from a central star by varying either
the amplitude of the central star's light in the focal plane
\citep{traub,kuchner03,debes04}.  

All of these designs in theory can
reach the contrast ratio necessary for imaging a planetary companion, however
most of these concepts are just starting to be tested in the lab or on a real 
telescope
where other concerns arise.  Each
design also has its own set of drawbacks.  Pupils or image plane
masks that require 
transmissive functions require 
precision to very high levels (10$^{-10}$) and are suceptible to
degradation in a space environment \citep{ks03}.  Designs that use
binary masks have less restrictive tolerances but must be fabricated
precisely.  The specific advantages of each idea cannot be 
determined until they are actually built and tested in such a way as to 
simulate real observing conditions. 

Apodization through binary shaped apertures that have either completely
transmissive or completely opaque openings represent one promising class
 of techniques.  An intuitive description of how the choice 
of a shaped aperture affects the resulting PSF is in order.
The entrance aperture can be described as a contour
$y\prime=\pm(1/2)C(x\prime)$, where $|x\prime| < x\prime_M$ the maximum horizontal extent of the aperture.  An
example of a familiar contour for an entrance aperture is 
$C(x\prime)=\sqrt{x\prime_M-4x\prime^2}$, the equation for a circular aperture.  The resulting
point spread function (PSF) as predicted by scalar Fraunhoffer diffraction 
theory is given by the 2-D Fourier transform of the aperture:
\begin{equation}
\label{eqn:axieta}
{\cal A}(\xi\prime,\eta\prime)=\int^{x\prime_M}_{-x\prime_M} \int^{\frac{1}{2}C(x\prime)}_{-\frac{1}{2}C(x\prime)} \exp(2\pi k i\xi\prime x\prime)\exp(2\pi k i\eta\prime y\prime)\mbox{d}x\prime \mbox{d}y\prime
\end{equation}
where $k=2x\prime_M/\lambda$ and we follow the notation of \citet{jacquinot64}.  We then transform this
 equation into the reduced coordinates $x=x\prime/2x\prime_M,
 y=y\prime/2x\prime_M, \xi=2\xi\prime x\prime_M/\lambda$, and $\eta=2\eta\prime x\prime_M/\lambda$.  The equation then becomes:
\begin{equation}
\label{eqn:axietared}
{\cal A}(\xi,\eta)=\int^{\frac{1}{2}}_{-\frac{1}{2}} \int^{\frac{1}{2}C(x)}_{-\frac{1}{2}C(x)} \exp(2\pi i\xi x)\exp(2\pi i\eta y)\mbox{d}x \mbox{d}y.
\end{equation}

This two dimensional
problem can be simplified to a one dimensional Fourier transform provided the 
PSF is restricted to a particular $\eta=\eta_o$:
\begin{equation}
\label{eqn:oned}
{\cal A}(\xi,\eta_o)=\int^{\frac{1}{2}}_{-\frac{1}{2}} \frac{\sin(\pi \eta_o C(x))}{\pi \eta_o} \exp(2\pi i\xi x) \mbox{d}x
\end{equation}
By sampling $\eta$ to the desired precision one can build up a 2-D PSF by
performing a 1-D Fast Fourier Transform (FFT) \citep{spie03}.  In this
way the PSF can be determined with more 
accuracy and less time than with traditional 2-D FFTs.  There exists
the special case of the diffraction pattern along the $\eta=0$ axis,
which even more greatly simplifies the expression by becoming a simple
Fourier Transform of $C(x)$:
\begin{equation}
\label{eqn:azero}
{\cal A}(\xi,0)=\int^{\frac{1}{2}}_{-\frac{1}{2}}C(x)\exp(2\pi i\xi x)\mbox{d}x
\end{equation}

Binary shaped apertures where $C(x)$ is described
as a truncated Gaussian function represent a 
promising design for high contrast
imaging and was suggested for extrasolar planet searches
by \citet{spergel01}.  However, the idea has potential uses 
on ground based telescopes as well for more modest goals, such as
high contrast imaging surveys that cannot afford to take the overhead
of aligning coronagraphic masks or who observe multiple targets in one 
field of view \citep{ge02}.  In this case the gaussian contour
is 
\begin{equation}
\label{eqn:cx1}
C(x)  =  a \left\{ \exp \left[-\left(2 \alpha x\right)^{2}\right] -\exp \left(-\alpha^{2}\right) \right\}.
\end{equation}
Since the Fourier transform of a
gaussian function is another gaussian, 
the amplitude of the diffraction pattern $A(\xi, \eta)$ along
the $\xi$ axis decreases like e$^{-\xi^2}$, which we denote the high contrast
 axis.  The intensity
distribution in the imaging plane is given by $|A A^{\star}|$.  
The ratio $z=I(\xi,\eta)/I(0,0)$ gives the contrast relative to the
peak intensity of the diffraction pattern.  The variables $a$ and $\alpha$
 are free parameters that can be
used to optimize the aperture for depth of contrast, the angle from the central
object at which high contrast starts, and the azimuthal area of high contrast.
It should be noted that based on this definition, changing $\alpha$ subtly 
changes the height of the contour since the maximum height of the contour
is $\propto 1-\exp(-\alpha^2)$.  However, in terms of the range of $\alpha$ that we
test and are interested in ($\alpha=2-4$) 
for realistic observations, this is a negligible
effect on the order of a few percent.

The final peak intensity $I(0,0)$ measured on a detector is proportional to 
the total area of the aperture squared.  In the reduced coordinates that we use, a 
square, fully transmissive aperture gives the maximum intensity, which we 
normalize to one \citep{jacquinot64}.  In this normalization, a circular aperture gives rise to a
 maximum intensity of $\pi^2/16$.  Determining the peak intensity for a GAPM
then requires summing the total area of the given aperture and squaring it:
\begin{equation}
I(0,0)=|\int^{\frac{1}{2}}_{-\frac{1}{2}} C(x)\mbox{d}x|^2.
\end{equation}
Solving this for Equation \ref{eqn:cx1}, gives $I(0,0)= (a \sqrt{\pi} 
 \mbox{erf}(1/2)/(2 \alpha)-a\exp(-\alpha^2))^2$.  We can compare this to a circular aperture.  For
 an $\alpha=2.7, a=1$ GAPM, this corresponds to a peak intensity 3\% of a full
circular aperture.  A Lyot stop that undersizes the circular aperture by 1/2
gives a peak 6.25\% of the original.  Light gathering power is proportional
to the area, and so for the same examples, the Lyot coronagraph would have
25\% throughput and the GAPM 17\%.

\begin{figure}
   \plotone{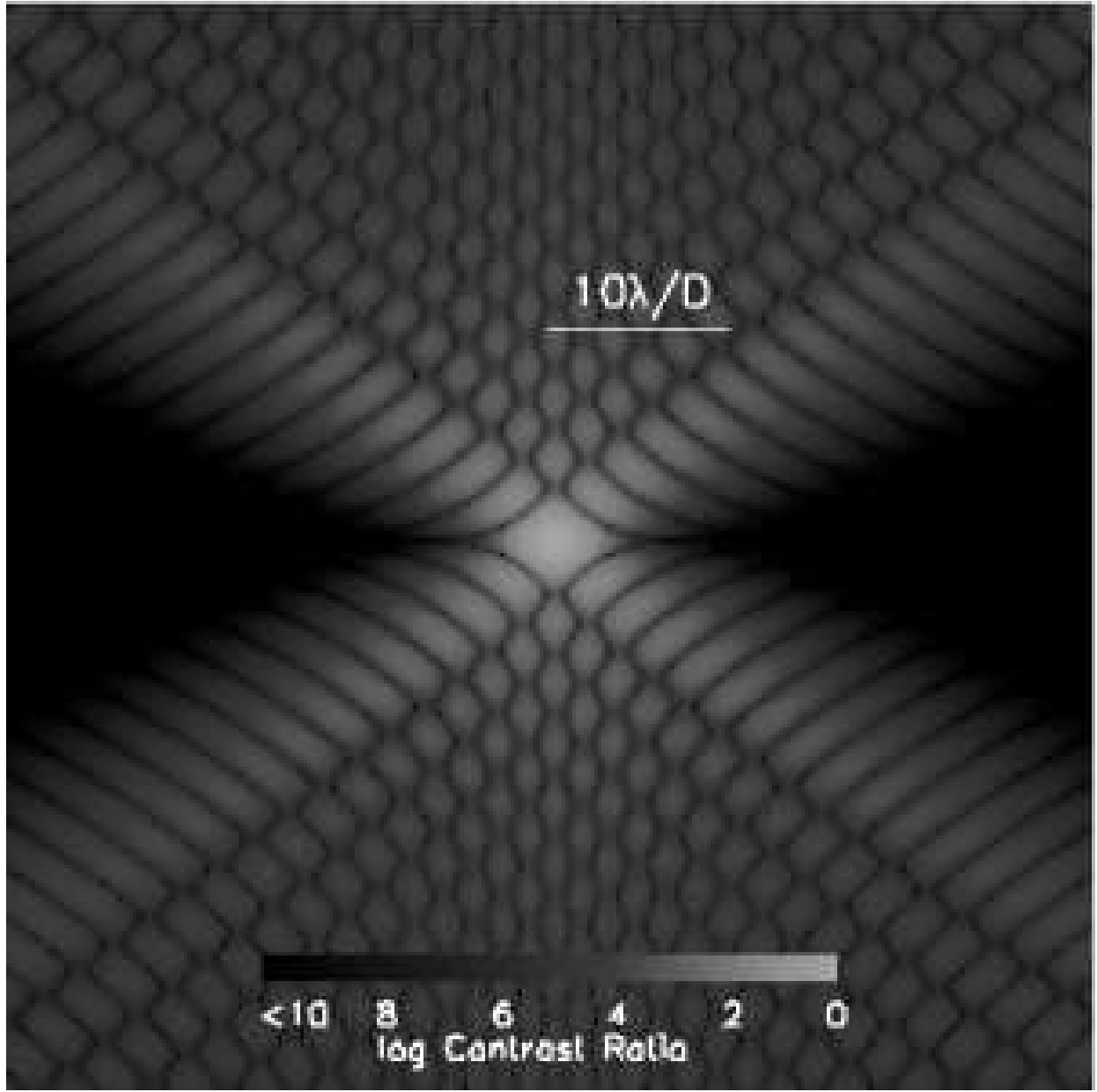}
   \caption
{ \label{fig:monopsf} The resulting monochromatic 
diffraction pattern for a single aperture GAPM with $a=1.$ and $\alpha=2.5$.
  The image is logarithmically scaled.
}
   \end{figure}

By substituting Equation \ref{eqn:cx1} into Equation \ref{eqn:oned}, the PSF 
can be determined.  Figure \ref{fig:monopsf} shows the resulting PSF for 
an aperture with $a=.6$ and $\alpha=2.9$, a typical pattern
from a Gaussian contour.  Since the contour is not azimuthally
symmetric neither is the PSF, with a region that has
high contrast, which we denote the high contrast region (HCR).  High
contrast is deepest along the $\xi$ axis.
  
If one is designing a mask, it would be useful to have several first 
order estimates of the dependance of $a$ and $\alpha$ on the resulting 
diffraction pattern.  We start with the contrast along the
axis of highest contrast and an approximate separation in $\xi$ where high 
contrast starts.  The exact solution of Equation 
\ref{eqn:azero} is given by
\begin{equation}
\label{eqn:azeroex}
{\cal A}(\xi,0)=\frac{\sqrt{\pi}}{2\alpha}\exp\left[\-\left(\pi\xi/2\alpha
\right)^2\right]\star\mbox{sinc}(\xi)-2\exp(-\alpha^2)\mbox{sinc}(\xi),
\end{equation} 
 where $\mbox{sinc}(\xi)=\sin(\pi \xi)/\pi\xi$.  This is difficult to manipulate analytically, but can be approximated to
first order by 
\begin{equation}
\label{eqn:azap}
A(\xi,0)=\frac{\sqrt{\pi}}{2\alpha} \exp[-( \pi\xi / 2 \alpha)^2] -
\exp[- \alpha^2]\mbox{sinc}(\xi), 
\end{equation} which ignores the truncation
of the exponential part of the function.  

The width of the PSF core
can then be estimated by determining the location of the first zero ($\xi_{fz}$) in Equation \ref{eqn:azap}.
After some manipulation one finds that
\begin{equation}
\label{eqn:fst0}
\xi_{fz}^2=\frac{4\alpha^2}{\pi^2}\left[\alpha^2-\mbox{ln}|\mbox{sinc}(\xi_{fz})|+\mbox{ln}|\frac{\sqrt{\pi}}{2 \alpha}|\right].
\end{equation}
One can solve this equation exactly through a simple recursion algorithm.  However, noting that $\mbox{ln}|\mbox{sinc}(\xi)|\sim3$ over the range of $\xi$ we
are interested in ($\xi=3-10$), that $\mbox{ln}|\frac{\sqrt{\pi}}{2 \alpha}|\sim-1$ over the range of interesting $\alpha$ values ($\alpha=2-4$), and assuming
$\alpha^2\gg2$,we find that Equation \ref{eqn:fst0} is well approximated by:
\begin{equation}
\xi_{fz}=\frac{2\alpha^2}{\pi} \left(1+\frac{1}{\alpha^2} \right).
\end{equation}

The approximate
 contrast one can expect to achieve is then found by substituting 
$\xi_{fz}$ into Equation \ref{eqn:azap} and squaring the result.
  Finally, one can estimate
the angular coverage of the HCR by noticing the HCR is governed by
the tail of the Gaussian function, which transitions over at the contour's 
change of curvature.  By solving for the maximum angle where the change occurs on the aperture,
the resulting angle of high contrast in the imaging plane is given by
\begin{equation}
\cot \theta=\sqrt{\frac{2}{e}}a\alpha.
\end{equation}

By placing a mask into the pupil plane with a Gaussian aperture, one can
transform a traditional circular aperture telescope into one with a diffraction
pattern better suited for high contrast imaging.  Using a mask represents a
quick, efficient, and economical way to test this emerging imaging method to 
determine its advantages and tradeoffs and compare them to the performance of
other existing techniques.  More subtle phenomena that limit contrast can also
be studied and removed with a well-known system whose ideal performance and 
performance under non-ideal conditions can be easily modeled.  Scattered light
from microroughness or polarization effects can be more reliably studied and 
verified experimentally rather than with completely theoretical treatments.

We have endeavored to begin anwering the question of which design ultimately 
will be useful in the search for extrasolar planets, or which will be most
useful for other areas of astronomy where less stringent tolerances are present.
To that end we have designed, fabricated, and tested several GAPM designs for
use with the Penn State near-IR Imager and Spectrograph (PIRIS)\citep{ge03}.  In Section 
\ref{design} we explore what the best design for a telescope would be.  In 
Section \ref{fab} we briefly discuss the process of fabrication of the GAPMs,
while in \ref{test} we discuss the various tests we performed in the lab and
on the ground at the Mt. Wilson 100$^{\prime\prime}$ telescope.  
Finally in Section \ref{concl} we discuss
what role GAPMs have for future high contrast imaging.
  
\section{Designing a GAPM for current telescopes}
\label{design}
The idealized design of a single gaussian aperture in practice cannot be used
on current telescopes due to their circular secondary obstructions and the 
presence of the support structure.  These two additions serve to modify the 
resulting diffraction pattern and destroy the advantages of the single 
aperture.  Therefore, a new design that avoids or minimizes their effect is 
necessary to retain high contrast.  There are two possible solutions: multiple
 apertures
that avoid the structure completely, or a way of blocking the structure without
changing the diffraction pattern in the imaging plane, such as with another
gaussian curve.

\subsection{Multiple Apertures}

By taking a contour over a sub-aperture of width $h D$ where $D$ is the 
diameter of the telescope pupil, one reproduces a similar diffraction pattern
as obtained from the contour described in Equation \ref{eqn:cx1} with the
exception that $\xi$ and $\eta$ must be rescaled by dividing by $h$.
  Placing multiple subapertures on the pupil
as the convolution of the base subaperture with a sum of N $\delta$-functions
in the desired positions of the subapertures.  The Fourier transform of that
convolution becomes:
\begin{equation}
\label{eqn:multi}
{\cal A}^\prime(\xi,\eta)={\cal A}(\xi,\eta)\sum^N_{k=1} \exp(2\pi i x_{k} \xi)
\exp(2\pi i y_{k} \eta).
\end{equation}

As mentioned in \citet{spergel01} and \citet{kasdin1}, one can also have 
multiple apertures with asymmetric sides as long as they are reflected about
the $y$-axis.  In this case the contour can be described on the full pupil
in the original reduced coordinates of Equation \ref{eqn:axieta} as
\begin{eqnarray}
\label{eqn:asym}
C_{1}\left(\frac{x}{h}\right) & = & h C\left(\frac{x}{h}\right)+2y_{o} \nonumber \\
C_{2}\left(\frac{x}{h}\right) & = & - \frac{hb}{a}C\left(\frac{x}{h}\right)+2y_{o}
\end{eqnarray}
where $y_{o}$ is a constant vertical offset from the $y$-axis and $a$ is not
equal to $b$.  The resulting diffraction pattern is the superposition of the
PSF from the 
two contours.  Either approach can yield a grouping of apertures that 
completely avoids the support structures and maximizes the possible throughput.
The potential tradeoff is a widening of the PSF core, as well as a new peak
that is multiplied by a factor of $N h^4$.

\subsection{Minimizing Support Structure}
Another option for avoiding the secondary of a telescope is to create an opaque
secondary gaussian curve that blocks the circular secondary mirror.  The result in the focal plane
would be the superposition of PSFs, which follows naturally from the 
Babinet principle.  One way of implementing this design is to have a contour
similar to Equation \ref{eqn:cx1} but with a second contour with a height $b$
where $b < a$:
\begin{eqnarray}
\label{eqn:gausssec}
C_{1}(x) & = & a C(x) \nonumber \\
C_{2}(x) & = & - b C(x).
\end{eqnarray}
  The resulting PSF retains the higher resolution of the telescope, though
angular coverage may be less than what could be achieved with multiple apertures.

In this case it is important to estimate the effect of not avoiding the support
structure.  The level of contamination by the support vanes for
the secondary mirror can be estimated by deriving the diffraction pattern
for a slit with the same proportions of the vanes and using Babinet's 
principle.  A vane of width $w$ along the $x$-axis
and length $l$ in the $y$-axis has an amplitude of 
\begin{equation}
\label{eqn:vane}
{\cal A}_{s}(\xi,\eta)=w l \mbox{sinc} ( \xi w) \mbox{sinc}( \eta l).
\end{equation} 
The vane will be brightest along the $\xi$ axis.  Ideally, support vanes
should be rotated 45$^\circ$ with respect to the HCR, so that the diffraction
spikes can be masked by the lower contrast regions of the PSF. 
${\cal A}_s$ can be compared to ${\cal A}(0)$. If ${\cal A}_{s} \ll {\cal A}(0)$,
 and 
the secondary is not so large that the loss in throughput is great, this
method may be preferable. As an example we look at spider vanes that have a
width of $\sim10^{-3}D$, which corresponds to a contrast of $10^{-8}$ at a 
distance of $5\lambda/D$ assuming that the two vanes are oriented 45$^\circ$
to the axis of interest.  Clearly this places a fundamental limit on the width
of any support structure (or gaps in a multi-mirror design) for an extrasolar
planet search.  Taking the limiting contrast to be $10^{-10}$ at 5$\lambda/D$,
the size limit is $10^{-4}D$.

\subsection{Final Design of the Prototype}  
\begin{figure}
   \plotone{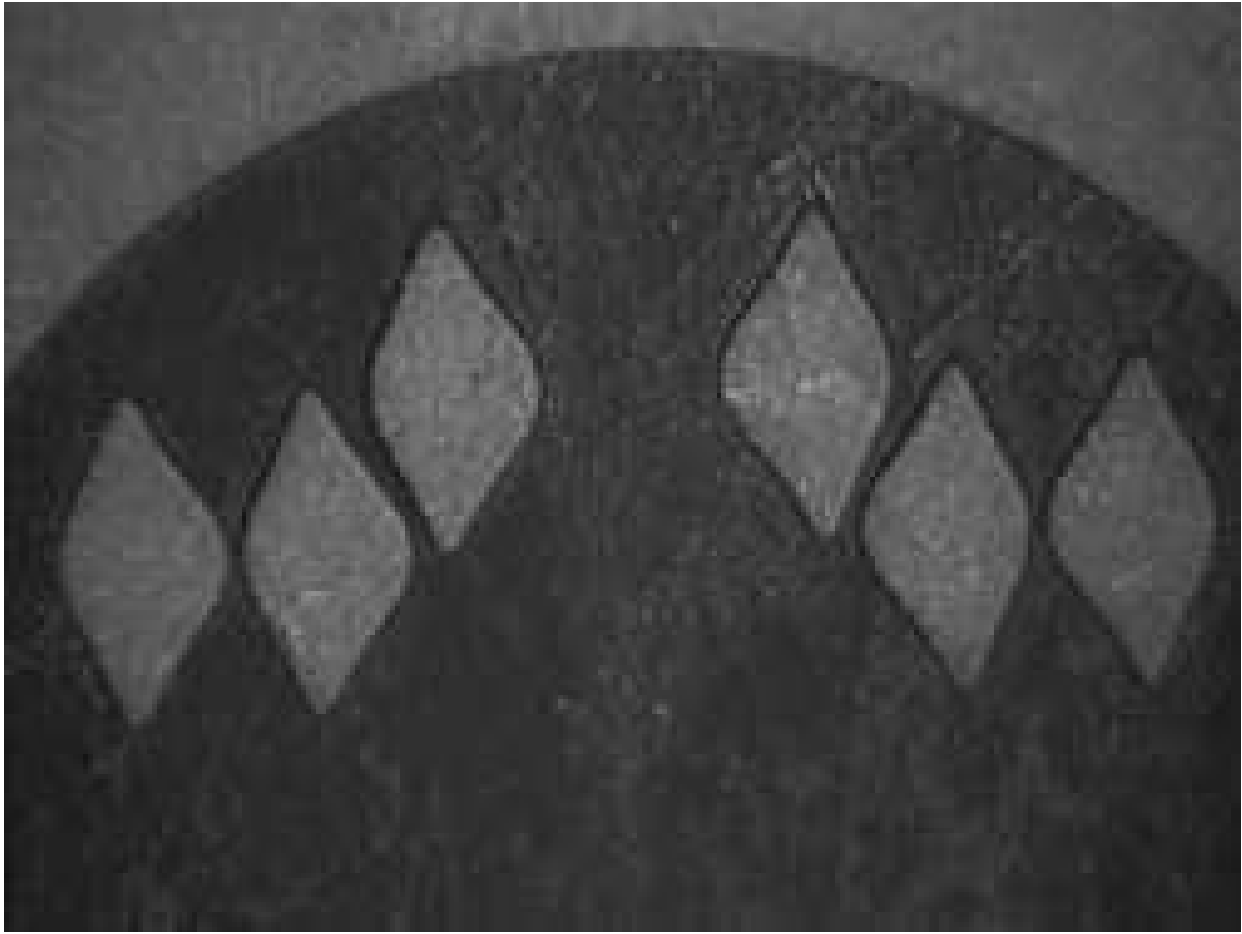}
   \caption{ \label{fig:realmask} The top half of the fabricated mask used for observing on the 
Mt. Wilson telescope at 5x optical magnification. 
 The twelve apertures allow $\sim$25\% throughput, while avoiding the support
 structure of the telescope.  This picture shows the 
imperfections in the fabrication process.
}
   \end{figure}

\begin{figure}
   \plotone{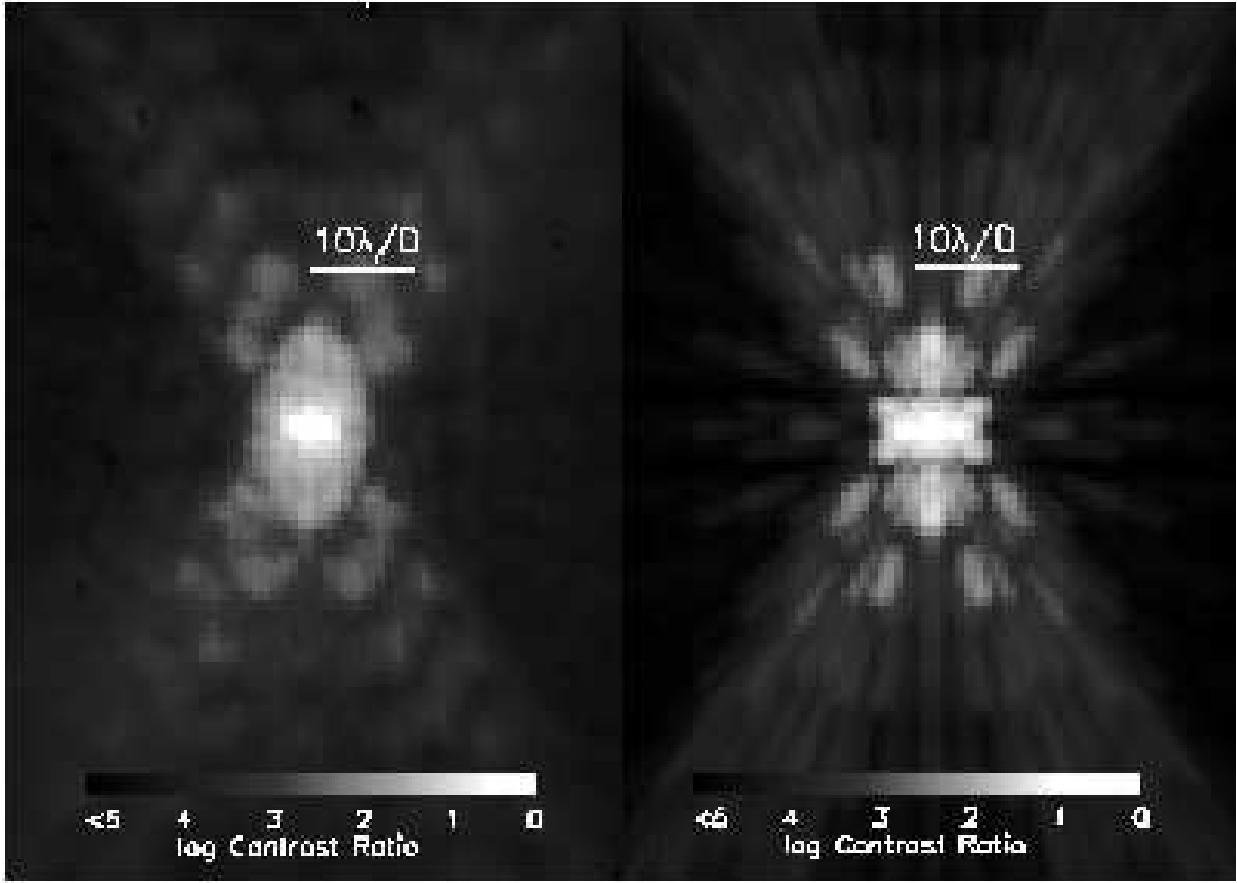}
   \caption
{ \label{fig:psfcomp} Comparison of the theoretical J band diffraction pattern
(right) compared with lab results with PIRIS (left).  Both are on the same 
logarithmic scale.  The model used included the imperfect shape of the mask 
that was fabricated and was integrated over the transmission function of a typical J filter.}
\end{figure}

We designed a mask to be used at the Mt. Wilson 100$^{\prime\prime}$ telescope
for preliminary observations, as well as a
single aperture design for testing in the lab.
  The diameter of the secondary at Mt. Wilson is $\sim$30\%
the diameter of the primary and the width of the spider vanes are $\sim$.25\%
of the diameter.  We decided to completely avoid the support structure for the
initial prototype to lower the risk of the PSF being contaminated by 
misalignments of the pupil mask.  For a final design we decided to try a 
variation of what was proposed by \citet{spergel01}, using a contour
based on Equation \ref{eqn:asym} by placing 3 sub-apertures
in each quadrant of the mask to maximize throughput to about
16\%.  Table \ref{tab:params}
shows the parameters that we used for the two masks as well as the positioning
of the apertures for the Mt. Wilson design.  Figure \ref{fig:realmask} shows
what the final design looked like.  Figure \ref{fig:psfcomp} shows a
comparison between a J band image taken with the mask and a theoretical PSF modeled by taking digital images of the apertures at high magnification and taking
a 2D FFT with IDL.  Since the spatial scale of the PSF determined in Equation \ref{eqn:axietared}
is scalable with wavelength, one can build a multi-wavelength PSF by adding
the scaled PSF in wavelength bins
 together and multiplying by the transmission of the particular filter used.

\section{Fabricating a GAPM}
\label{fab}
Once a design was chosen the masks were fabricated.  We chose to fabricate
the Mt. Wilson and lab
designs with Photo-Chemical Machining
(PCM).  This technique has been used to produce masks to block thermal radiation
 from telescope structures for near-IR observing and for creating Lyot stops \citep{mccarthy01}.
 
The process of PCM, also called Photo-Etching or Photo-Chemical Milling, 
involves using a thin metal sheet that is coated with a light sensitive polymer.  A UV photo imaging tool is used to imprint the desired design on the
sheet.  It is then developed much like film and chemically etched by an
aqueous solution of ferric chloride (FeCl$_3$).

Several masks are present in the PIRIS camera mainly for the traditional Lyot
coronagraphic modes.  They were fabricated by Newcut, Inc. (Newark, NJ).  
For the GAPMs we submitted CAD designs based on the simulations performed
to Newcut and they fabricated the masks.  A sheet of 25-50 masks 
with a diameter of $\sim$ 4 mm were 
fabricated at very low cost within a few weeks.  When they were delivered
they were photographed on an optical telescope with 5x and 50x magnification.

This technique can provide the basic shape we need, but 
has difficulty preserving the exact shape of the design in the smallest 
regions.  The edges of the gaussians on the mask were truncated well before 
they 
mathmatically would be.  Variations on the order of 10 $\mu m$ are 
also observed
in the masks.  Both of these imperfections can degrade contrast, which is 
discussed further in Section \ref{test}.  These imperfections are likely caused
by the photo printing as well as the chemical etching.  For instance, the
corner truncation and width variation can be caused by underexposure of the 
light sensitive polymer.  The rugged edge can be caused by non-uniform chemical
etching.  Since the etching is isotropic, changes in the physical and 
chemical conditions of the etchant can cause local irregularities.

\section{Testing the Prototype}
\label{test}
\subsection{First Light and Lab tests}
We placed the Mt. Wilson design on the Penn State Near-IR Imager and 
Spectrograph (PIRIS) and ran tests both in the lab and on the 
100$^{\prime\prime}$ Mt. Wilson
telescope.  We used the prototype as part of a survey for faint companions
 around nearby solar type 
stars \citep{chakraborty02,debes02}.

Further details are outlined in \citet{debes02}.  We used the prototype GAPM
to study two stars, $\epsilon$ Eridani and $\mu$ Her A.  The mask allowed 
confirmation of a common
proper motion companion to $\mu$ Her A and placed limits on any possible
companions
around $\epsilon$ Eridani down to the level of a 38$M_{jup}$ brown dwarf.  This
corresponds to a performance an order of magnitude better than adaptive 
optics alone and 2 times worse than our Lyot coronagraph performance without 
blocking the central star's light 
\citep{debes02}.

Lab testing was also performed in the J and H bands on the Mt. Wilson
design.  The setup involved taking an
incandescent lamp and simulating a point source to sample the PSF generated by
the different masks.  An optical fiber took light
from an incandescent lamp where it passed through a micro objective and a 
pinhole.  The light then was collimated by a collimator
achromat.  After the collimator it was focused onto the slit wheel 
aperture by an image achromat.
The image achromat 
also forms
an exit pupil, $\sim$1.9 m away from the focal plane, mimicking the Mt. Wilson 
100$^{\prime\prime}$
exit pupil.  On the slit wheel we placed our focal plane coronagraphic masks.
The light then travels through the camera optics of PIRIS where it is read
by the 256 $\times$ 256 PICNIC array.

\begin{figure}
\plotone{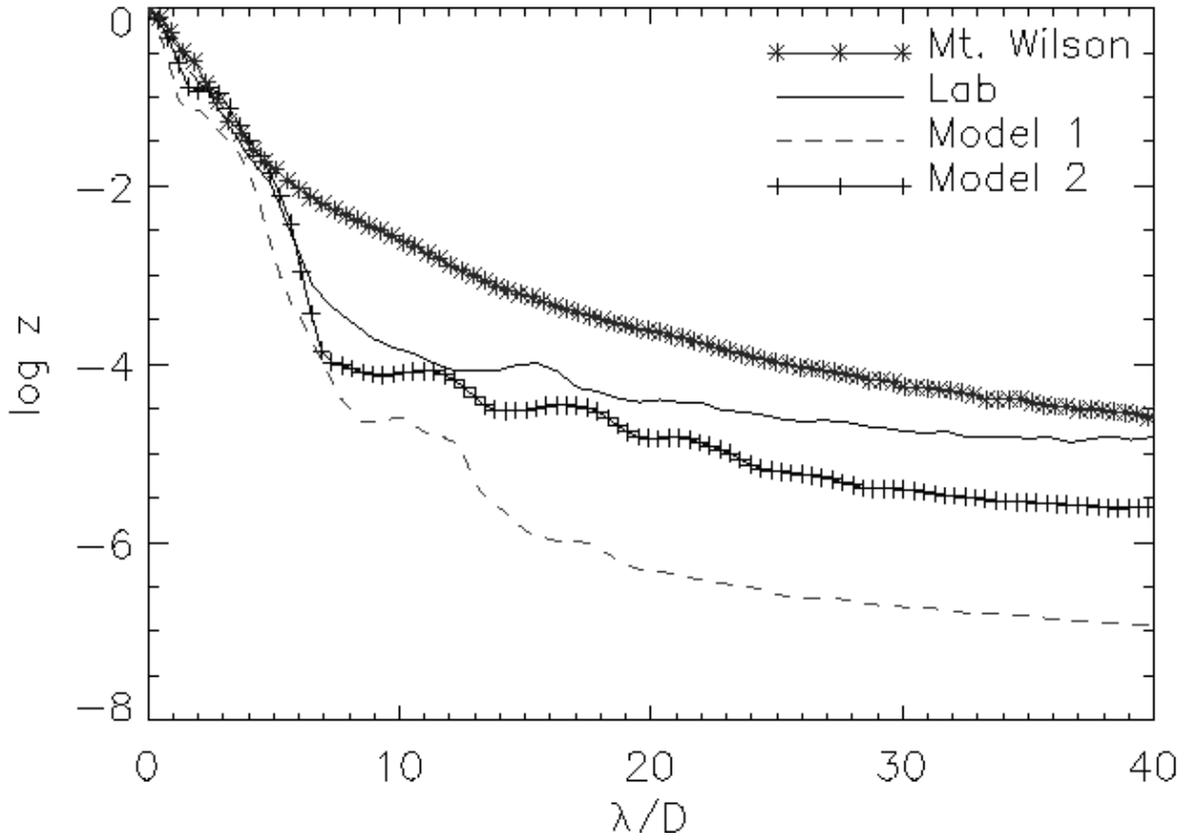} 
   \caption
{ \label{fig:datacomp}  Azimuthally averaged
PSF profiles for the GAPM multiaperture design tested at Mt. Wilson, along with
results of a lab test in the J band and two different models.  Model 1 
represents a model with perfect apertures, while Model 2 is the simulated
results based on the actual shape of the apertures.
}
\end{figure}

Figure \ref{fig:datacomp} shows an azimuthally averaged comparison 
between the data taken at Mt. Wilson,
lab tests in the J band, and two theoretical multiwavelength PSFs as a 
function of $\lambda_o/D$, where $\lambda_o$ represents the central wavelength
of the filter used.
  For the lab and models this corresponds to using a J 
filter whereas on Mt. Wilson all observations were done in the K band.  
Azimuthal averaging was performed over $\pm 30^{\circ}$
from the high contrast axis on both sides.
One simulation, called model 1, represents a completely ideal situation where
the mask is perfectly created and no wavefront errors exist.
The second simulation, model 2, takes
the observed shape of the masks under magnification as the apertures and 
neglects other errors.  One can see that the theoretical
simulation of the observed shape matches the lab data quite well, off by less 
than an order of magnitude close to the center.  The observed shape errors also
degrade the contrast achievable by the idealized design.  Finally the effect
of the atmosphere is present in the Mt. Wilson data.  From our observations,
the seeing was $\sim1^{\prime\prime}$, giving an $r_o\sim46$~cm at 2.2 $\mu m$
.  We estimate
that our Strehl ratio for most of the observations was lower than expected,
$\sim$.1.  At this level of correction the halo from the atmosphere severely
degrades the contrast to $\sim10^{-2}$.

From images of the pupil optics of PIRIS, we observed low amounts of 
light leakage and thermal emission.  This slightly degrades contrast, which would explain the 
slight difference between our models and the observed lab PSF.  

\begin{figure}
   \plotone{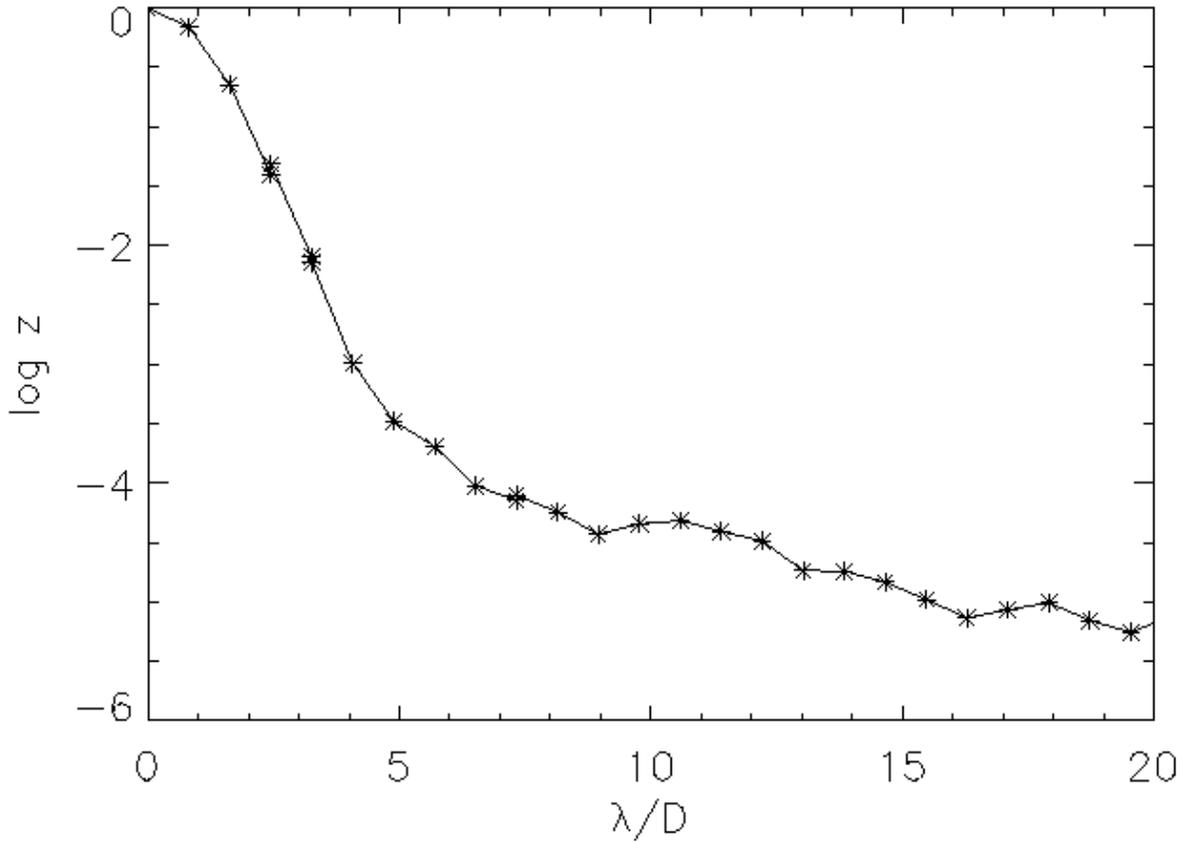}
   \caption {\label{fig:labcomp} Azimuthally averaged PSF of the single 
aperture design in the J band.  We achieved $4\times10^{-5}$ contrast at 10$\lambda/D$.  Truncation of the mask during fabrication is the main degradation of
contrast with what would be possible theoretically.}
\end{figure}

The single aperture GAPM was also tested in the J band in the lab.  Figure \ref{fig:labcomp} shows the contrast achieved with an azimuthal average over 
$\pm20^{\circ}$ with respect to the high contrast axis.  The truncation
present in the fabrication also severely degrades the contrast possible with
this mask.  We find that we can achieve a contrast of $\sim10^{-5}$ at 10$\lambda/D$.

\subsection{Modeling the Degradation in Contrast}
Figure \ref{fig:realmask} shows one of the greatest limitations to the current prototype, which is the truncation of the thinnest parts of the mask.  This is due
to errors in the fabrication process as well as a lack of resolution to 
reproduce the true contour.  We can model this effect by taking 
images of the apertures and digitizing them into model apertures, which has 
been shown above to be effective in Figures \ref{fig:psfcomp} and \ref{fig:datacomp}.  

There is also a way to express this analytically, by slightly changing the 
contour of the aperture:
\begin{equation}
\label{eqn:cxtrunc}
C(x)  =  a \left\{ \exp \left[-\left(2\alpha x\right)^{2}\right] -\exp(2\beta x) 
\exp \left(-\alpha^{2}\right) \right\},
\end{equation}
where $\beta$ is a free parameter that can be determined empirically.  The 
effect of this change raises the floor of contrast by increasing the strength
of the second term in Equation \ref{eqn:cxtrunc}.

\begin{figure}
   \plotone{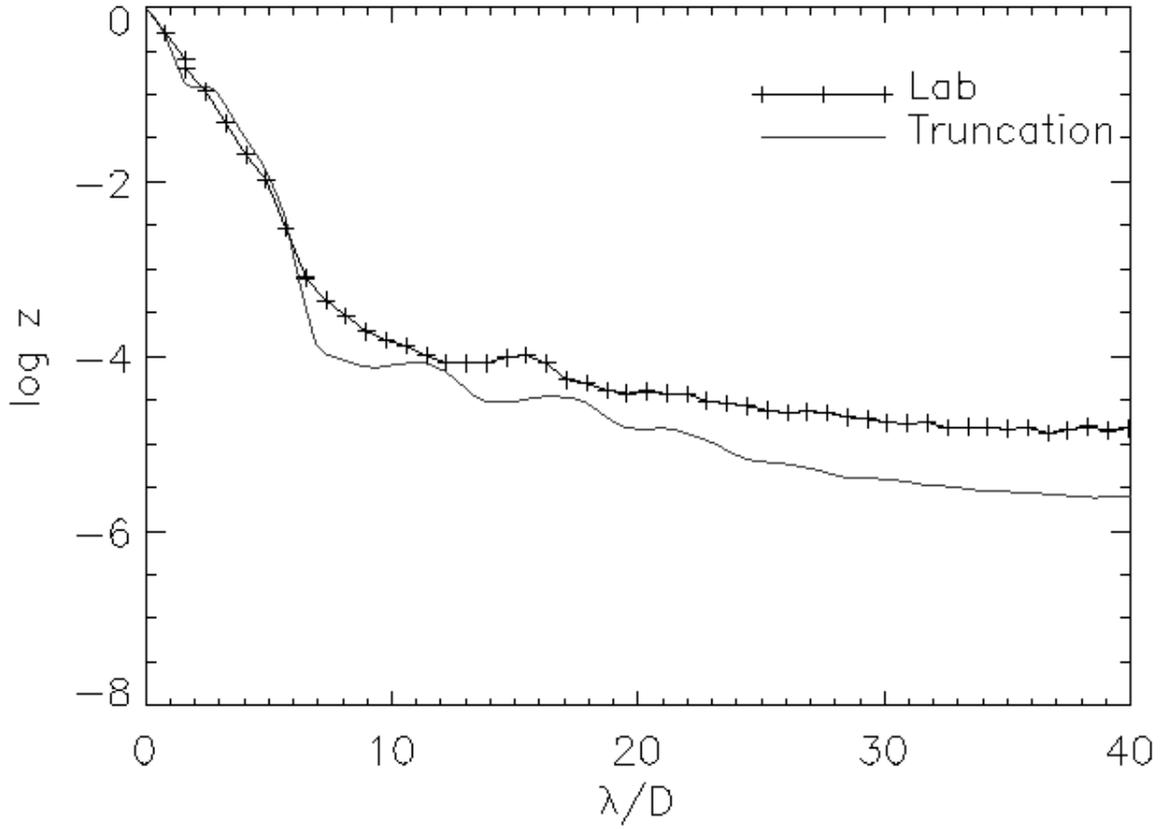}
   \caption
{ \label{fig:truncomp}
A comparison between the J band PSF and a model based on the truncation of the
Gaussian contour's wings described in Equation \ref{eqn:cxtrunc}.  For the model
we used $\beta$=13 and the parameters listed in Table \ref{tab:params} for the
Mt. Wilson design.  Both the model and the lab data were azimuthally averaged 
for 20$^\circ$ about the high contrast axis.
}
   \end{figure}

Figure \ref{fig:truncomp} shows a comparison between the observed multi-wavelength PSF
of our prototype Mt. Wilson design and a model based on Equation \ref{eqn:cxtrunc} both
azimuthally averaged over $20^{\circ}$ around the high contrast axis.  A
$\beta$=13 matches the observed PSF and the models based on the fabricated
apertures quite well.

Another potential problem beside truncation of the edges is the tolerance for
accuracy in the mask itself.  We can estimate the effect by considering either
a square opening or obstruction of side $h$ depending on the kind of error that occurred.
This extra square will create its own diffraction pattern that will constructively add to the PSF.  Using the same equation as in Equation \ref{eqn:vane}, we
can estimate the intensity due to the total number of errors present for a particular distance, $(\sim N_{err}h^2/\pi\xi)^2$, where $N_{err}$ is the
total number of errors present.  In the case of errors
due to the resolution of a fabrication method one can base the estimate on the
fact that the number of errors will be $\sim h^{-1}$ where we assume the size 
of the error is no bigger than the resolution of the machine.  This leads to
the contrast degrading as $h^2$.  In order to reach a contrast of $10^{-10}$ 
at $4\lambda/D$,
errors are restricted to $< 10^{-4}D_p$, where $D_p$ is the pupil mask
diameter.  For a pupil mask with a diameter of 
4~mm, this corresponds to a resolution of .4 $\micron$. 

\subsection{Lessons for AO observers}
The lessons that can be gained from the AO observations can be put to use for
future work.  It is clear that a high order of correction is needed for the 
full advantage of the GAPM to be utilized.  Ideally one would choose an inner
working distance (IWD) that is less than the region that is fully corrected 
(to the level of the residual halo)
by the AO system.  This angle, $\theta_{c} = N_{act}\lambda/2D$, depends on the
number of actuators across the diameter of the primary \citep{sivar01}.
For Mt. Wilson, this area extends out to $8\lambda/D$ so a design with 
an IWD of closer to $4\lambda/D$ or less would be ideal.  

Ground based 
observations are also limited by the contrast ratio of the uncorrected halo
to the core ($z_{\mbox{halo}}$).  The FWHM of the halo is approximately the 
uncorrected seeing, which typically is $> \theta_c$.  In that case a good 
estimate of the best contrast achievable is simply the ratio of the peak 
intensity of the uncorrected halo to the peak intensity of the corrected image
core \citep{hardy}:
\begin{equation}   
z_{\mbox{\small halo}}=\frac{1-S}{1+\left(\frac{D}{r_o}\right)^2},
\end{equation}
where $S$ is the Strehl Ratio, $r_o$ is the Fried Parameter of atmospheric
turbulence, and $D$ is the 
diameter of the telescope.  \citet{roberts02} has reported $S$ values of $\sim$.2 for the 
AEOS telescope in the visible (.55 $\mu m$) for an $r_o\sim12$ cm, giving an approximate contrast of 10$^{-3}$.  In this case, larger 
telescopes gain an advantage in contrast given the same seeing conditions and
ability to attain a certain Strehl ratio.  Higher order adaptive optics will
need to correct to very high levels (S$\sim.9$) to successfully achieve a high 
contrast with a GAPM.  However, using a more modest $\alpha$ with a more modest
level of high contrast may present a useful alternative to a coronagraph
for groundbased searches with AO, for example in relatively dense young 
clusters or for faint companion searches to white dwarfs.  

\section{Conclusions}
\label{concl}
We have performed several simulations, lab tests, and telescope observations 
with GAPMs in order to better understand
the interplay between theory and the reality of observations.  GAPMs alone 
provide an improvement over a simple circular aperture for quick high contrast
imaging.  They are very sensitive to an accurate reproduction of shape and thus
need accuracies that require nanofabrication techniques such as e-beam lithography, similar to what has been used for 
notch-filter masks \citep{debes04}.  We have commissioned some
precise GAPMs with accuracies on the order of .25~$\micron$ which should 
lead to a more accurate shape reproduction and a minimization of errors. 
Precisely fabricating these masks can potentially improve performance to 
the ideal limit for a mask provided it is above the scattered light
limit of the telescope, bringing it in line with Lyot coronagraphs
 of comparable throughput.  Demonstration of these masks in conjunction with
 an adaptive optics system 
could present a workable example of a quick way to survey for faint companions
without needing to
incur the overhead cost of precise alignment behind a coronagraphic image mask.

\acknowledgements

The authors would like to acknowledge D. McCarthy for loaning part of the optics for PIRIS, R. Brown (LPL) for
the PICNIC array, C. Ftaclas for coronagraphic masks, and A. Kutyrev for
filters.  Several important comments from the anonymous referee
were particularly useful as well as discussions with D. Spergel and M. Kuchner which were crucial in
our understanding of gaussian apertures and band limited masks.
We would also
like to thank the invaluable help of the
the Mt. Wilson staff, L. Engel for design help with the GAPM, Abhijit 
Charkraborty, and Dan McDavitt for help providing figures.

 This
work was supported by NASA with grants NAG5-10617, NAG5-11427 by NSF with grants NSF ASTRO-0138235 by Ball Aerospace, Inc. as well as the
Penn State Eberly College of Science.  J.D acknowledges funding by a NASA GSRP fellowship under grant NGT5-119.

\bibliographystyle{apj}
\bibliography{spie}

\begin{thebibliography}{20}
\expandafter\ifx\csname natexlab\endcsname\relax\def\natexlab#1{#1}\fi

\bibitem[{{Brown} \& {Burrows}(1990)}]{brown91}
{Brown}, R.~A. \& {Burrows}, C.~J. 1990, Icarus, 87, 484

\bibitem[{{Chakraborty} {et~al.}(2002){Chakraborty}, {Ge}, \&
  {Debes}}]{chakraborty02}
{Chakraborty}, A., {Ge}, J., \& {Debes}, J.~H. 2002, \aj, 124, 1127

\bibitem[{{Debes} {et~al.}(2002){Debes}, {Ge}, \& {Chakraborty}}]{debes02}
{Debes}, J.~H., {Ge}, J., \& {Chakraborty}, A. 2002, \apjl, 572, L165

\bibitem[{{Debes}(2004)}]{debes04}
{Debes}, J.~H. e.~a. 2004, \apj, (in press)

\bibitem[{{Ge}(2002)}]{ge02}
{Ge}, J. e.~a. 2002, in Future Research Direction and Visions for Astronomy.
  Edited by Dressler, Alan M. Proceedings of the SPIE, Volume 4835, pp. 87-97
  (2002)., 87--97

\bibitem[{{Ge}(2003)}]{ge03}
{Ge}, J. e.~a. 2003, in Instrument Design and Performance for Optical/Infrared
  Ground-based Telescopes. Edited by Iye, Masanori; Moorwood, Alan F. M.
  Proceedings of the SPIE, Volume 4841, pp. 1503-1514 (2003)., 1503--1514

\bibitem[{{Hardy}(1998)}]{hardy}
{Hardy}, J.~W. 1998, {Adaptive optics for astronomical telescopes} (Oxford
  University Press)

\bibitem[{{Jacquinot} \& {Roizen-Dossier}(1964)}]{jacquinot64}
{Jacquinot}, P. \& {Roizen-Dossier}, B. 1964, Progress in Optics, 3, 31

\bibitem[{{Kasdin}(2003)}]{kasdin1}
{Kasdin}, N.~J. e.~a. 2003, \apj, 582, 1147

\bibitem[{{Kuchner} \& {Spergel}(2003{\natexlab{a}})}]{ks03}
{Kuchner}, M.~J. \& {Spergel}, D.~N. 2003{\natexlab{a}}, in ASP Conference
  Series 294, Scientific Frontiers In Research In Extrasolar Planets, ed. D.
  Deming \& S. Seager (2003)

\bibitem[{{Kuchner} \& {Spergel}(2003{\natexlab{b}})}]{kuchner03}
{Kuchner}, M.~J. \& {Spergel}, D.~N. 2003{\natexlab{b}}, \apj, 594, 617

\bibitem[{{Kuchner} \& {Traub}(2002)}]{traub}
{Kuchner}, M.~J. \& {Traub}, W.~A. 2002, \apj, 570, 900

\bibitem[{{Lieber}(2003)}]{spie03}
{Lieber}, M.~D. e.~a. 2003, in Techniques and Instrumentation for Detection of
  Exoplanets. Edited by Coulter, Daniel R. Proceedings of the SPIE, Volume
  5170, pp. 66-78 (2003)., 66--78

\bibitem[{{Lyot}(1939)}]{lyot}
{Lyot}, B. 1939, \mnras, 99, 580

\bibitem[{{Malbet}(1996)}]{malbet96}
{Malbet}, F. 1996, \aaps, 115, 161

\bibitem[{{McCarthy}(2001)}]{mccarthy01}
{McCarthy}, D.~W. e.~a. 2001, \pasp, 113, 353

\bibitem[{{Nisenson} \& {Papaliolios}(2001)}]{nisenson01}
{Nisenson}, P. \& {Papaliolios}, C. 2001, \apjl, 548, L201

\bibitem[{{Roberts} \& {Neyman}(2002)}]{roberts02}
{Roberts}, L.~C. \& {Neyman}, C.~R. 2002, \pasp, 114, 1260

\bibitem[{{Sivaramakrishnan}(2001)}]{sivar01}
{Sivaramakrishnan}, A. e.~a. 2001, \apj, 552, 397

\bibitem[{{Spergel}(2001)}]{spergel01}
{Spergel}, D.~N. 2001, astro-ph/0101142

\end{thebibliography}

\begin{deluxetable}{lcccccc}
\tablecolumns{6}
\tablewidth{0pc}
\tablecaption{\label{tab:params}Table of the parameters and positioning
used for the Mt. Wilson and Lab apertures.  These paramters are described
in relation to Equations \ref{eqn:cx1} and \ref{eqn:asym}.}
\tablehead{
\colhead{GAPM Design} & \colhead{$\alpha$} & \colhead{$a$} & \colhead{$b$}
& \colhead{$x_k$} & \colhead{$y_k$} & \colhead{$h$}}
\startdata
Mt. Wilson & 2.7 & 0.23 & 0.33 & ($\pm$0.3125,$\pm$0.1625,$\pm$0.1625) &
(0.1,0.2125.,0.3875) & 0.5 \\
Lab & 2.7 & 1. & - & 0 & 0 & 0.8 \\
\enddata
\end{deluxetable}

\clearpage


\end{document}